\documentclass[aps,pra,showpacs,twocolumn,superscriptaddress]{revtex4-1}

\usepackage{graphicx}% Include figure files
\usepackage{amsmath,amssymb}

\usepackage{color}% Include color
\usepackage{comment}% Include comment

\begin{document}

\title{
%Magnetic properties of ultracold fermions in multilayered Lieb lattices
%Flat band magnetism in the multilayered Lieb optical lattice
Flat-band ferromagnetism in the multilayer Lieb optical lattice
}

\author{Kazuto Noda}
\email{noda.kazuto@lab.ntt.co.jp}
\affiliation{NTT Basic Research Laboratories, NTT Corporation, Atsugi 243-0198, Japan}
\affiliation{JST, CREST, Chiyoda-ku, Tokyo 102-0075, Japan}
\author{Kensuke Inaba}
\affiliation{NTT Basic Research Laboratories, NTT Corporation, Atsugi 243-0198, Japan}
\affiliation{JST, CREST, Chiyoda-ku, Tokyo 102-0075, Japan}
\author{Makoto Yamashita}
\affiliation{NTT Basic Research Laboratories, NTT Corporation, Atsugi 243-0198, Japan}
\affiliation{JST, CREST, Chiyoda-ku, Tokyo 102-0075, Japan}

\begin{abstract}
We theoretically study magnetic properties of two-component cold fermions in half-filled multilayer Lieb optical lattices, i.e., two, three, and several layers, using the dynamical mean-field theory. We clarify that the magnetic properties of this system become quite different depending on whether the number of layers is odd or even. In odd-number-th layers in an odd-number-layer system, finite magnetization emerges even with an infinitesimal interaction. This is a striking feature of the flat-band ferromagnetic state in multilayer systems as a consequence of the Lieb theorem. In contrast, in even-number layers, magnetization develops from zero on a finite interaction. These different magnetic behaviours are triggered by the flat bands in the local density of states and become identical in the limit of the infinite-layer (i.e., three-dimensional) system. We also address how interlayer hopping affects the magnetization process. Further, we point out that layer magnetization, which is a population imbalance between up and down atoms on a layer, can be employed to detect the emergence of the flat-band ferromagnetic state without addressing sublattice magnetization.
\end{abstract}

% pacs
%05.30.Fk   Fermion systems and electron gas (see also 71.10.-w Theories and models of many-electron systems; see also 67.10.Db Fermion degeneracy in quantum fluids)
%37.10.Jk   Atoms in optical lattices
%71.10.Fd   Lattice fermion models (Hubbard model, etc.)
\maketitle

%%%%%%%%%%%%%%%%%%%%%%%%%%%%%
\section{Introduction}
%%%%%%%%%%%%%%%%%%%%%%%%%%%%%
%{\color{green} %begin blue
Cold atoms loaded into an optical lattice have opened up a new field dedicated to the study of quantum magnetism, which has been a long-standing problem in condensed matter physics \cite{Bloch2008,Esslinger2010}.
According to the well-known Stoner criterion, the interaction strength and the band structure at the Fermi energy determine whether the magnetic transitions occur or not.
The key advantage of cold atoms over condensed matter is controllability of the interactions between atoms and the lattice geometry characterizing the energy band structure. %, which encourage us to regard this system as a quantum simulator of magnetism.
These advantages encourage us to regard this system as a quantum simulator of magnetism.
By controlling the interactions, the Mott transition of fermionic atoms has been successfully demonstrated \cite{Jordens2008,Schneider2008}, which is an important step in the investigation of quantum magnetism.
%Another big advantage is that the lattice geometry can be engineered with the recent experimental techniques for creating optical lattices.
In addition, recent experimental techniques allow us to create various complex lattices, such as honeycomb, kagom\'e, Lieb lattices \cite{Becker2010,Wirth2011,Soltan-Panahi2011,Jo2012,Tarruell2012,Uehlinger2013}.

The Lieb lattice is a prominent example of lattices that provide interesting topics related to magnetism.
Part of the energy band structure of this lattice is dispersionless, which is called a flat band.
When the flat band is at the Fermi energy, the magnetic transition can occur with infinitesimally small (positive) interactions because the density of states (DOS) of atoms at the Fermi energy is infinitely large.
This strong instability toward the magnetic phase transition can be easily understood from the Stoner criterion.
From another point of view, the occurrence of the magnetic transition of this lattice has been mathematically demonstrated by Lieb \cite{Lieb1989}, in what is called the Lieb theorem.
%{\color{blue}
%Note that these two understandings about the flat-band ferromagnetism are just two aspects to see the same. 
%The Stoner criterion is a viewpoint from a band picture, while the Lieb theorem is that from a mathematical picture.
%} % color
Although many theoretical studies on this flat-band magnetism have been performed \cite{Mielke1991,Tasaki1992,Mielke1993,Tasaki2008}, the experimental realization of this lattice has not yet been achieved in condensed matter.
Recently, Refs. \cite{Shen2010,Apaja2010} have theoretically proposed a laser configuration to construct a Lieb optical lattice, and the Kyoto University group has successfully achieved its construction \cite{Takahashi}.
%This proposal also stimulates further theoretical research, for instance, investigations of topological phases \cite{Weeks2010,Goldman2011,Zhao2012,Nita2013}.

Most of the previous studies on the Lieb lattice have been done in two dimensions \cite{Noda2009,Shen2010,Apaja2010,Weeks2010,Goldman2011,Zhao2012,Nita2013}, whereas experiments will be performed in a three-dimensional laser configuration.
A realistic Lieb optical lattice structure is a stack of two-dimensional lattices.
However, such a feature specific to cold atoms has not been well discussed.
This stimulates us to investigate how the interlayer correlation affects the magnetism.
%For example, the interlayer hopping changes the band structure, and thus the magnetic properties of the layered Lieb lattice will be different from the single layer two-dimensional lattice.
On the other hand, in condensed matter, the layered materials are now attracting much interest thanks to the experimental progress in this field; {\it e.g.,} hetero-structure materials of the correlated electron systems \cite{Ohtomo2002} and multilayer graphene with control of the number of layers \cite{CastroNeto2009}.
Various studies on such systems have clarified that multilayer systems contain rich physics beyond single-layer or bulk material \cite{Hwang2012}.
Layered optical lattices \cite{Uehlinger2013} can also be regarded as quantum simulators for investigating such a current topic in condensed matter.

In this study, we investigate the magnetic properties of two-component fermions in a multilayer Lieb lattice at half filling and at zero temperature.
We study the interlayer correlation effects in detail by systematically changing the number of layers.
We elucidate that the magnetic processes for the even and odd number of layers are completely different when the interaction is small.
This difference disappears when the number of layers is infinite, namely, at the three-dimensional limit.
We also discuss how interlayer hopping affects the magnetism by changing the magnitude of this parameter.
We find that this additional energy scale not included in the two-dimensional system characterizes whether or not an exotic crossover specific to the layered system occurs.
%the flatband  only in odd layers.
%} %end blue

%We also reveal that, both for odd and even layers, magnetization process in the small interaction region is determined by the flat band structure in local density of states (DOS).
%We demonstrate how these magnetization process are altered as the number of layer increases.

%{\color{blue}
Our paper is organized as follows.
In Sec. \ref{sec_model}, we derive the our model Hamiltonian from the experimental laser potential.
In Sec. \ref{sec_method}, we briefly explain our theoretical method.
In Sec. \ref{sec_res}, we discuss the magnetic properties in multilayer Lieb lattice systems.
In Sec. \ref{sec_con}, we briefly summarize our results.
%}

%%%%%%%%%%%%%%%%%%%%%%%%%%%%%
\section{Model}
\label{sec_model}
%%%%%%%%%%%%%%%%%%%%%%%%%%%%%
%%%%%%%%%%%%%%%%%%%%%%%%%%
\begin{figure}[t]
\begin{center}
\includegraphics[width=0.9\linewidth]{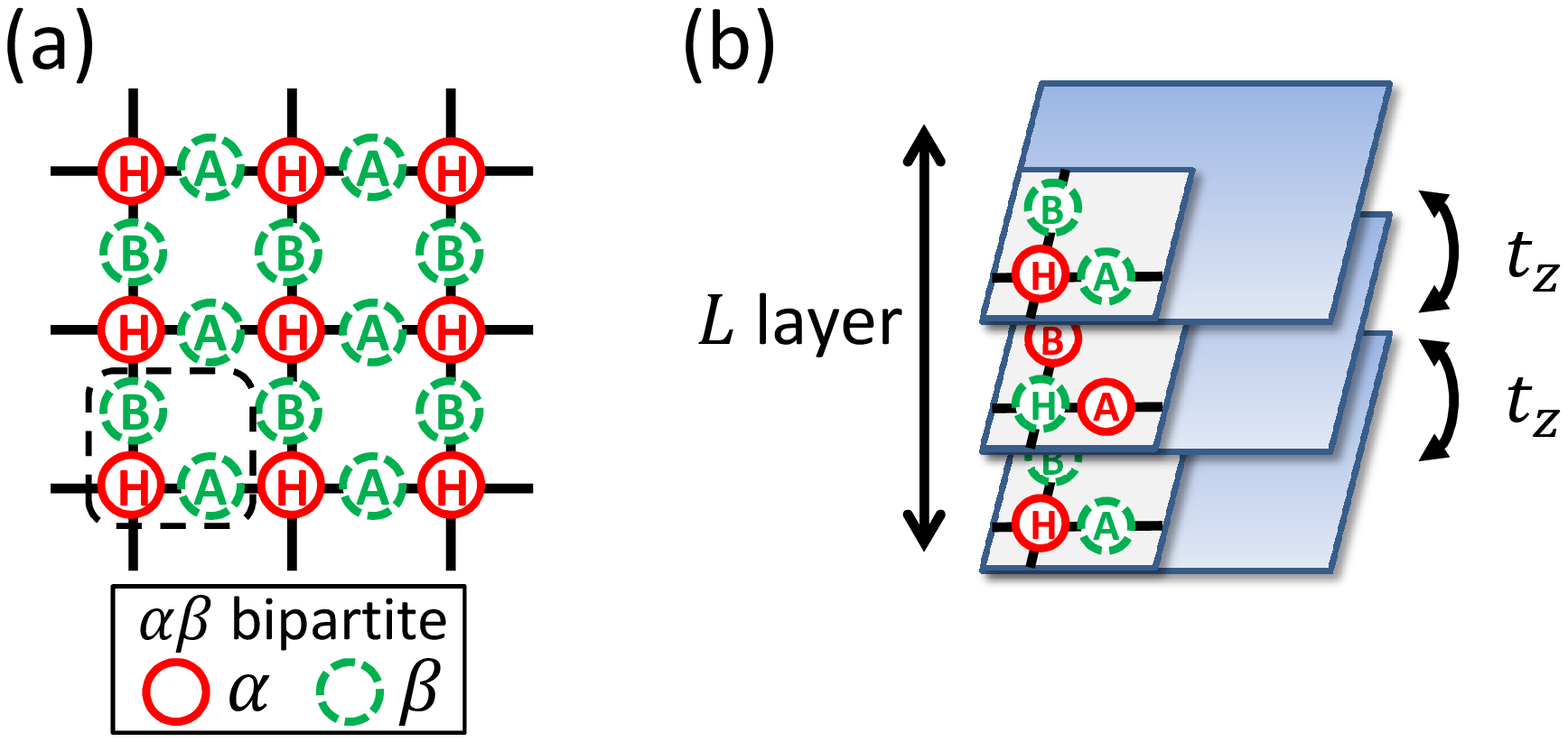}
\caption{(Color online) (a) The Lieb lattice is constructed by sites H, A, and B. Because this lattice is $\alpha\beta$ bipartite, site H (A, B) belongs to sublattice $\alpha$($\beta$). The dashed line shows the unit cell. (b) Schematic picture of the multilayer system. $t_z$ represents interlayer hopping. Note that site H does not always belong to sublattice $\alpha$, while the multilayer system is also $\alpha \beta$ bipartite.}
%\includegraphics[clip,width=0.9\linewidth]{gg1dos.eps}
%\caption{(Color online). (a) The Lieb lattice, where a dashed line shows the unit celsl. (b) A schematic picture of the multilayered system. $t_z$ represents interlayer hopping.}
\label{fig_scpic}
\end{center}
\end{figure}
%%%%%%%%%%%%%%%%%%%%%%%%%%%%

%{\color{green}
We start by explaining the structure of the multilayer Lieb lattices. %; and related to this, we also discuss the Lieb theorem in these lattices.
Figure\,\ref{fig_scpic}\,(a) depicts the two-dimensional Lieb lattice, where the sites at hubs and spokes are labeled $H$ and $A$ or $B$, respectively.
The dotted square in Fig.\,\ref{fig_scpic}\,(a) represents the unit cell.
Figure\,\ref{fig_scpic}\,(b) schematically describes the multilayer lattice with the number of layers $L$ of $3$, where the sites in the unit cell are only shown.
As illustrated in these figures, the multilayer Lieb lattices are bipartite, where all sites can be classified into sublattice $\alpha$ or $\beta$ that have inter-sublattice ($\alpha$-$\beta$) connections only while no intra-sublattice ($\alpha$-$\alpha$ or $\beta$-$\beta$) connections. % between $\alpha$-$\alpha$ or $\beta$-$\beta$.
These sublattices are represented as (red) solid and (green) dashed circles in Fig.\,\ref{fig_scpic}.
%{\color{blue} 
%
In general, the magnetism in such bipartite structural lattices can be discussed on the basis of the Lieb theorem.
%} % end of blue

The Lieb theorem states that, in bipartite lattices, two-component fermions show a finite magnetization with an infinitesimal repulsive interaction at half filling and at zero temperature \cite{Lieb1989}.
The magnitude of the total magnetization is given by $M_{\rm tot}=1/2(N_{\alpha}-N_{\beta})$, where $N_\alpha$ ($N_\beta$) is the total number of the sites in sublattice $\alpha (\beta)$. 
Figure\,\ref{fig_scpic} clearly shows that, for the multilayer Lieb lattice, sublattice $\alpha$ ($\beta$) includes sites $H$ ($A$ and $B$) in the odd-number-th layers and sites $A$ and $B$ ($H$) in the even-number-th layers.
The Lieb theorem predicts that the values of $M_{\rm tot}$ per unit cell are $0.5$ for odd $L$ and $0$ for even $L$ in the multilayer Lieb lattices.
%This indicates that $M_{\rm tot}$ per unit cell is $0.5$ for odd $L$, and $0$ for even $L$, which are the Lieb theorem in the multilayered Lieb lattices.
%} % end of green
%{\color{blue}
Note that this theorem does not provide any concrete information about local quantities, such as local magnetization, which can be calculated with the aid of numerical methods.
%} % end of blue

Our present study is primary concerned about how the interlayer correlations induce the interesting phenomena beyond the predictions of the Lieb theorem. We next show that layered Lieb optical lattices can be derived straightforwardly from the ingenious configuration of lattice lasers.
As discussed in Refs.\,\cite{Shen2010,Apaja2010}, the single-layer (two-dimensional) Lieb lattice can be created by the potential $V_{\rm Lieb}(x,y)=V[ \cos(2\pi x/a)+\cos(2\pi y/a)+\cos(\pi x/a)+\cos(\pi y/a)+1/2\sin\big(\pi (x+y)/a\big)+1/2\sin\big(\pi (x-y)/a\big) ]$, where $a$ is the lattice distance and  $V$ is the lattice potential depths.
The implementation of this complex potential requires several lasers with different wavelengths, {\it e.g.}, $2a, 4a$, and $2\sqrt{2}a$ \cite{Shen2010,Apaja2010}.
On the other hand,  we usually apply the additional potential along the $z$ direction to confine cold atoms three-dimensionally in the real experiments. 
As a result, a total potential is
\begin{eqnarray}
\label{eq_pot}
V(x,y,z)
&=&
V_{\rm Lieb}(x,y)+V_z \cos(2\pi z/a_z),
\end{eqnarray}
where  $a_z$ and $V_z$ are the distance and depth, respectively, for the lattice potential along the $z$ direction.

The model Hamiltonian of cold atoms in the potential in Eq. (\ref{eq_pot}) is naturally given by Hubbard Hamiltonian on the layered Lieb lattice,  ${\cal H}$, when $V$ and $V_z$ are much larger than the recoil energy of atoms $E_r$ \cite{Bloch2008}:
\begin{eqnarray}
{\cal H}&=&
{\cal H}_{\text{Lieb}}+{\cal H}_{z}+{\cal H}_{U},
\\
{\cal H}_{\text{Lieb}}&=&
- t \sum_{m \sigma} \sum_{\langle i,j \rangle} c_{mi\sigma}^{\dagger} c_{mj\sigma},
\\
{\cal H}_{z}
&=&
- t_z \sum_{i \sigma} \sum_{\langle m,m' \rangle} c_{mi\sigma}^{\dagger} c_{m'i\sigma},
\\
{\cal H}_{U}
&=&U \sum_m \sum_{i} n_{mi\uparrow}n_{mi\downarrow},
\end{eqnarray}
where $c_{mi\sigma} (c_{mi\sigma}^{\dagger})$ is an annihilation (creation) operator of an atom with spin $\sigma$ at site $i$ on the $m$-th layer, and the number operator is defined as $n_{mi\sigma}=c_{mi\sigma}^{\dagger}c_{mi\sigma}$.
The subscript $\langle i,j \rangle (\langle m,m' \rangle)$ denotes the summation over the nearest neighbor sites in the $xy$ plane ($z$ direction).
The potentials in Eq. (\ref{eq_pot}), $V_{\rm Lieb}(x,y)$ and $V_z \cos(2\pi z/a_z)$, determine the intralayer and interlayer hopping amplitude, $t$ and  $t_z$, respectively.
%The interaction strength $U$ can be determined from the scattering length $V, V_z, a$ and $a_z$.
We assume the present Hamiltonian to be uniform and neglect inhomogeneity due to the trapping potential, which will not change our results qualitatively and just modify them quantitatively.
In this paper, we only consider the system at half filling without an external field and at zero temperature as a first step.
We set $t=1.0$ as a unit of energy.

To clearly discuss the interlayer correlation effects on the flat-band magnetism, we investigate the $L$-layer Lieb lattices as shown in Fig.\,\ref{fig_scpic}\,(b) by systematically changing the number of layers as $L=2, 3, \cdots$.
Here, we set the periodic boundary conditions along the $x$ and $y$ directions, while an open boundary condition for the $z$ direction, which is a natural extension of our previous study for $L=1$ \cite{Noda2009}.
We also investigate the system with $L=\infty$, which corresponds to a three-dimensional layered Lieb lattice with periodic conditions for all three directions.
Then, we discuss the asymptotic behaviour from finite to infinite layers, which will show dimensional crossover from the two ($L=1$) to three dimensions ($L=\infty$) through quasi-three dimensions ($1<L<\infty$).

The two-dimensional Lieb lattice can also be realized in the limit of $t_z \to 0$.
Note that this limit is more realistic in experiments, which can be achieved by setting $V_z \gg V \gg E_r$.
However, we can  at most create an ensemble of the two-dimensional Lieb lattices, and the magnitude of $V_z$, which is proportional to the laser intensity, is limited for some practical reasons, such as a limitation on the laser power.
Therefore, to study the effects of the interlayer correlations on such an ensemble, we investigate the present model Hamiltonian by changing $t_z$ toward a small value.
Then, we discuss another dimensional crossover from the (quasi-)three dimensions ($t_z\not=0$) to the two dimensions ($t_z=0$).

%\textcolor{blue}{
The finite-$L$-layer systems can be implemented using the standard experimental techniques.
One of the methods is the selection of the layers. 
For instance, the atoms except for those in the selected layers will be taken away from the lattice with the radio-frequency knife.
%(or some spectroscopic ways) 
Another method without the loss of atoms is as follows. 
We can create an ensemble of the finite-$L$-layer lattices by superimposing extra lattice potentials with commensurate wavelengths along the $z$ direction. 
For example, the potential $V_z \cos^2(\pi z/a_z)+V'_z \cos^2(2\pi z/a_z) [+V''_z \cos^2(3\pi z/a_z)]$ can provide the two-layer [three-layer] Lieb lattice. 
These simple methods allow us to systematically study the interlayer correlation effects as will be discussed in Sec.\,\ref{sec_res}.
%}

%%%%%%%%%%%%%%%%%%%%%%%%%%%%%
\section{Method}
\label{sec_method}
%%%%%%%%%%%%%%%%%%%%%%%%%%%%%
%{\color{blue}
We use the dynamical mean field theory (DMFT) \cite{Metzner1989,Georges1996} to investigate the magnetic properties of multilayer Lieb lattices.
In the DMFT framework, each site in the original lattice problem is mapped onto an impurity with hybridization from a dynamical heat bath that effectively describes a connection to surrounding sites.
By solving this effective impurity problem in a self-consistent manner, we can precisely deal with local correlation effects, which are essentials for various quantum many-body phenomena, such as the Mott transitions, magnetism, and superconductivity.
%It is known that the DMFT framework is exact for the infinite dimensional systems and good for the higher, {\it e.g.}, three and more, dimensions.  %(three dimensional)
In fact, the DMFT used for layered systems \cite{Potthoff1999,Okamoto2004} has successfully demonstrated various phenomena experimentally observed in layered matter, such as a metal-insulator transition at the interface of a heterostructure of transition metal oxides \cite{Ohtomo2002}.
%{\color{blue}
We comment about our application of DMFT to the Lieb lattice, which consists of different coordination number sites. 
With the infinitesimal interaction, our DMFT calculations for magnetization, detailed in the next section, are consistent with analytical results described in Appendix. 
This indicates that our DMFT treatment is efficient to describe the flat-band ferromagnetism, which is one of the main topics of our paper.
%}

% in the interface of a heterostructure of transition metal oxides, such as, the emergence of metallic phase in the interface of a Mott and band insulating heterostructure \cite{Ohtomo2002}

To obtain actual DMFT solutions, the effective impurity problem should be solved with the aid of some numerical methods.
%To solve the effective impurity model,
The solver used in this study is the numerical renormalization group (NRG) \cite{Wilson1975,Bulla2008}.
This non-perturbative method can provide numerically exact solutions of the impurity problem in terms of thermodynamic properties at low temperatures.
In addition, it can very accurately calculate the low-energy spectral properties at around the Fermi energy.
The spectral anomaly of the flat band just at the Fermi energy is the origin of the instability toward magnetic ordering.
Thus, the NRG solver can capture the essence of the flat-band magnetism \cite{Noda2009}.
%In addition, the NRG can calculate thermodynamic quantities and spectral functions in wide parameter regions.
% We can calculate thermodynamic quantities and spectral functions in wide parameter regions.
%The lattice Green's function is obtained via the self-consistent solution of this impurity problem.
%Whereas the DMFT become exact in infinite dimensions,

Before detailing our DMFT approach, we should mention the unit cells of the present lattices.
A unit cell for the two-dimensional Lieb lattice comprises the three sites shown in Fig.\,\ref{fig_scpic}\,(a), and those for the $L$-layer lattices consist of $3\times L$ sites.
For $L=\infty$, the unit cell reduces to $6$ sites because of the two-site-period translational symmetry along the $z$ direction resulting from the antiferromagnetic ordering along this direction.
Note that the ordering pattern we take into account in this study is easily determined from the fact that the layered Lieb lattice is bipartite.
The size of the unit cells determines the matrix dimensions of the Green's function mentioned below.
% and the number of the (unique) effective impurity problems.
%the antiferromagnetic order along $z$-direction results from the fact that the layered Lieb lattice is bipartite.
%Even for finite $L$, a mirror symmetry along the $z=0$ plane allows us to identify some unit cells.

We explain our DMFT calculations.
Here, we self-consistently calculate $G_{m\gamma\sigma}(\omega)$ the local Green's function of atoms with spin $\sigma(=\uparrow, \downarrow$) at the site $\gamma(=H, A, B$) on the $m$-th layer ($m=1, 2, \cdots, L$):
\begin{equation}
\label{eq_Gloc}
G_{m\gamma\sigma}(\omega)=\left[\int d{\bf k}  \hat{G}_{{\bf k},\sigma}(\omega)\right]_{m\gamma,m\gamma},
\end{equation}
%%%%%%%%%%%%%%%%
where ${\bf k}$ is a wave vector ${\bf k}=(k_x, k_y)$, and
$\hat{G}_{{\bf k},\sigma}(\omega)$ is the following matrix with a dimension of $3L \times 3L$:
\begin{eqnarray}
\label{eq_Gk}
\hat{G}_{{\bf k},\sigma}(\omega)=
 \left [
\omega\hat{I}
-\hat{H}_\text{Lieb}({\bf k})-\hat{H}_{z}-\hat{\Sigma}_{\sigma}(\omega)
\right]^{-1},
\end{eqnarray}
where
$\hat{H}_{z}$ and $\hat{H}_\text{Lieb}({\bf k})$ are the matrix representation of the Hamiltonians ${\cal H}_{z}$ and the Fourier transform of ${\cal H}_{\rm Lieb}$, respectively, and $\hat{I}$ is the identity matrix.
The self-energy matrix $\hat{\Sigma}_{\sigma}(\omega)$ is diagonal within the DMFT framework, and it is now defined as
\begin{equation}
\hat{\Sigma}_{\sigma}(\omega)
=
{\text{diag}}(\Sigma_{1H\sigma}(\omega),\Sigma_{1A\sigma}(\omega),\Sigma_{1B\sigma}(\omega),\dots),
\nonumber
\end{equation}
where ${\Sigma}_{m\gamma\sigma}(\omega)$ is the self-energy of an atom with spin $\sigma$ at site $\gamma$ on the $m$-th layer, which can be obtained as mentioned below.
%{\color{blue}
Furthermore we set these self-energies to be zero at the first step of iterations. 
%} % color
%from the effective impurity problems as mentioned below.
On the other hand, from the Dyson equation, we obtain a cavity Green's function \cite{Georges1996}: % ${\cal G}_{m\gamma\sigma}$:
\begin{equation}
\label{eq_Gcav}
{\cal G}_{m\gamma\sigma}(\omega)=[G_{m\gamma\sigma}(\omega)^{-1}+{\Sigma}_{m\gamma\sigma}(\omega)]^{-1},
\end{equation}
which characterizes the dynamical heat bath connected to an effective impurity that corresponds to an atom with spin $\sigma$ at site $\gamma$ on the $m$-th layer in the original lattice.
By solving such effective impurity problems with the NRG, we obtain self-energy ${\Sigma}_{m\gamma\sigma}(\omega)$, which allows us to again calculate local Green's function $G_{m\gamma\sigma}(\omega)$ in Eqs. (\ref{eq_Gloc}) and (\ref{eq_Gk}).
This again yields a new ${\cal G}_{m\gamma\sigma}(\omega)$, which leads to a new ${\Sigma}_{m\gamma\sigma}(\omega)$.
We perform these calculations repeatedly until convergence.
%\textcolor{blue}{
Note that each layer is described by the corresponding three effective impurity problems. 
Our DMFT formulation for multilayers, which have several sites in the unit cell, is an straightforward extension of the Hubbard model with the antiferromagnetic order in bipartite lattices \cite{Georges1996,Noda2009}.
%}

For $L=\infty$, we can straightforwardly extend the above treatment by replacing wave vector ${\bf k}$ in Eqs. (\ref{eq_Gloc}) and (\ref{eq_Gk}) with ${\bf k}=(k_x, k_y, k_z)$  and replacing $\hat{H}_z$ in Eq. (\ref{eq_Gk}) with the Fourier transform $\hat{H}_z(\bf{k})$. Here all matrices have a dimension of $6\times 6$.

The self-consistently obtained $G_{m\gamma\sigma}(\omega)$ and NRG solutions provide us with various dynamical and thermodynamical quantities. % as shown in Sec.\,\ref{sec_res}.
We calculate the magnetization $M_{m\gamma}=\big(\langle n_{m\gamma\uparrow} \rangle - \langle n_{m\gamma\downarrow} \rangle\big)/2$, which are the order parameter of the magnetic transition. 
%{\color{blue}
In experiments, we can obtain these magnetizations from site- and spin-resolved observations of the number density, e.g., which can be achieved by microscopy and/or spectroscopy.
%}
%$\int d\omega [G_{m\gamma\uparrow}(\omega) -  G_{m\gamma\downarrow}(\omega)]$, and
We also calculate the local DOS of the atoms, $\rho_{m\gamma\sigma}(\omega)= -(1/\pi) {\rm Im}G_{m\gamma\sigma}(\omega+i \eta)$, which describes the single particle excitation spectra.
The local DOS clearly explains various phenomena.
For instance, the opening of a spectral gap clarifies the appearance of a metal-insulator transition.
%{\color{blue}
This excitation spectra were successfully measured in cold Fermi gases (without lattice potential) by JILA group \cite{Stewart2008}.
We can expect that this technique will be applicable to lattice systems.
%{\color{blue}
When we obtain the complete $k$-resolved excitation spectrum in a lattice, we can construct local DOS that we will discuss in the next section. 
%}
%{\color{blue}
Furthermore, we note here that the above quantities $M_{m\gamma}$ and $\rho_{m\gamma\sigma}(\omega)$ satisfy the symmetric relations with respect to $m$ and $\gamma$, reflecting a particle-hole symmetry on the AB bipartite lattice under the condition that filling is exactly half in the absence of an external field. 
These relations are broken when the system is away from at half filling or subject to any external fields, which is beyond the scope of our present study. 
%}
%} % end of blue

% with $\sigma$ spin at the $\gamma$ site on the $m$-th layer, which can be obtained by solving corresponding effective impurity models.
%We use a notation of diagonal components of the lattice Green's function as $G_{\gamma,\sigma}(\omega)=[\hat{G}_{\sigma}(\omega)]_{\gamma\gamma}$.]

%}%end color blue

%%%%%%%%%%%%%%%%%%%%%%%%%%%%%
\section{results}
\label{sec_res}
%%%%%%%%%%%%%%%%%%%%%%%%%%%%%

%%%%%%%%%%%%%%%%%%%%%%%%%%%%%
\subsection{Noninteracting DOS}
\label{sec_dos}
%%%%%%%%%%%%%%%%%%%%%%%%%%%%%
%%%%%%%%%%%%%%%%%%%%%%%%%%
\begin{figure}[tb]
\begin{center}
\includegraphics[clip,width=0.9\linewidth]{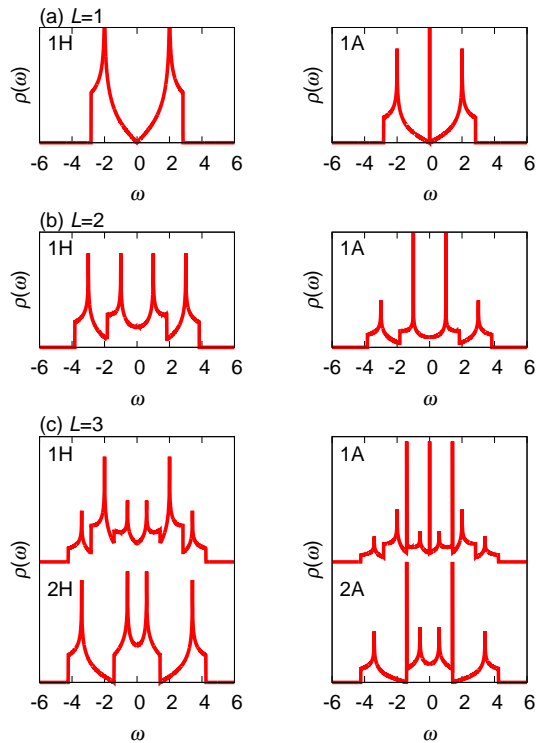}
\caption{(Color online) DOS for (a) $L=1$, (b) $L=2$, and (c) $L=3$ with $t_z=1.0$. "1A" means site A on the first layer. Note that the relation $\rho_{mA}(\omega)=\rho_{mB}(\omega)$ ($m=1,\ldots,L$) exists for all layers. Additionally, $\rho_{1\gamma}(\omega)=\rho_{2\gamma}(\omega)$ for $L=2$ and $\rho_{1\gamma}(\omega)=\rho_{3\gamma}(\omega)$ for $L=3$ ($\gamma=H,A,B$) exist.}
\label{fig_ggdos}
\end{center}
\end{figure}
%%%%%%%%%%%%%%%%%%%%%%%%%%%%

%{\color{green}
Before discussing magnetism based on the DMFT calculations, to see our way clearly,
we first provide the flat-band structures of noninteracting atoms in the multilayer Lieb lattice.
Figure \ref{fig_ggdos} shows $\rho_{m\gamma\sigma}(\omega)$ the DOS for noninteracting atoms at the site $\gamma(=H,A,B)$ on the $m$-th layer ($m=1,2,\cdots, L$) in the $L$-layer lattice for $L=1, 2,$ and $3$.
Note that the noninteracting DOS is independent of spin $\sigma$, and the mirror symmetries through $x=y$ and $z=0$ planes impose the relations $\rho_{mA\sigma}(\omega)=\rho_{mB\sigma}(\omega)$ and $\rho_{m\gamma\sigma}(\omega)=\rho_{L-m+1,\gamma\sigma}(\omega)$, respectively.
Here, we set $t_z=t=1.0$.

Figure\,\ref{fig_ggdos}\,(a) shows $\rho_{m\gamma\sigma}(\omega)$ for the two-dimensional Lieb lattice ($L=1$).
We find that the flat-band structure, namely, a delta function resulting from the dispersionless band structure, appears at Fermi energy $\omega=0$ in the DOS for site $A$, while no flat band appears for site $H$.
The flat band at the Fermi energy induces strong instability toward the magnetic ordering, and thus the ferromagnetism appears with the infinitesimal interaction $\delta U$, which can be understood from the well-known Stoner criterion.
Note that this mechanism of the flat-band ferromagnetism is consistent with the statement in the Lieb theorem as mentioned in Sec.\,\ref{sec_model}.

For multilayer lattices, interlayer hopping $t_z$ affects the flat-band structure in the DOS for site $A$, while it never moves the flat bands to the DOS for the site $H$.
As shown in Fig.\,\ref{fig_ggdos}\,(b), for $L=2$, two flat bands appear away from the Fermi energy.
Their spectral positions are located at $\omega = \pm t_z$, the estimations of which are detailed in Appendix.
For larger $L (> 2)$, such spectral features depend on layer number $m$.
Figure\,\ref{fig_ggdos}\,(c) shows that the first layer ($m=1$) in the three-layer lattice has three flat bands at $\omega=$ 0 and $\pm \sqrt{2}t_z$, while the second layer ($m=2$) has two flat bands at $\pm \sqrt{2}t_z$.
This difference in the flat-band structure is a key to the magnetism in multilayer lattices, as will be discussed in Secs.\,\ref{sec_odd} and \ref{sec_even}.

Generally, the DOS for the odd-number-th layers in odd-$L$-layer lattices have the flat band at $\omega=0$, while those for the even-number-th layers in odd-$L$-layer lattices and all of the layers in the even-$L$-layer lattices do not have the flat band at $\omega=0$.
Appendix explains this general important feature of the noninteracting DOS in details, and
section\,\ref{sec_L_dep} presents the DMFT calculations for general odd- and even-$L$-layer lattices.

For $L=\infty$, the spectral positions of the (infinite) flat bands become continuous, meaning that the flat-band structure becomes dispersive with the broadening of $W=4t_z$.
However, we find that the instability toward flat-band magnetism still remains in this limit.
This striking feature is discussed in Sec.\,\ref{sec_L_dep}.

A change in $t_z$ moves the spectral positions of flat bands as mentioned above.
We should note that some energy scales related to the flat-band positions  are important for understanding the quantitative properties of the magnetic phase transition or crossover in the present system.
This point is mainly discussed in Secs.\,\ref{sec_L_dep} and \ref{sec_tz_dep}.

%}% end green
% This is caused by the fact that the mirror symmetry of the $z$ direction, which does not break the particle-hole symmetry in the $xy$ plane, protects which layer the flat band on the Fermi energy appears in the odd layers.

%%%%%%%%%%%%%%%%%%%%%%%%%%%%%
\subsection{Three-layer system as a typical example of odd $L$}
\label{sec_odd}
%%%%%%%%%%%%%%%%%%%%%%%%%%%%%

%{\color{green}
%%%%%%%%%%%%%%%%%%%%%%%%%%
\begin{figure}[tb]
\begin{center}
\includegraphics[clip,width=0.9\linewidth]{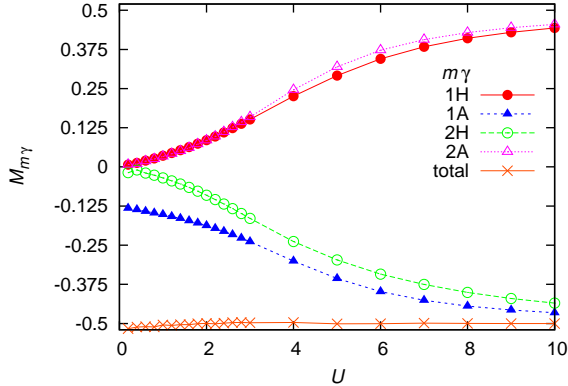}
\caption{(Color online) Sublattice magnetization of site $\gamma(=H,A,B)$ on $m(=1,2,3)$-th layer in the three-layer system ($L=3$) with $t_z=1.0$. Filled (blank) symbols denote the first (second) layer. Note that relations $M_{mA}=M_{mB}$ and $M_{1\gamma}=M_{3\gamma}$ exist.}
\label{fig_mag_3lay}
\end{center}
\end{figure}
%%%%%%%%%%%%%%%%%%%%%%%%%%%%

%%%%%%%%%%%%%%%%%%%%%%%%%%%%
\begin{figure}[tb]
\begin{center}
\includegraphics[clip,width=1.0\linewidth]{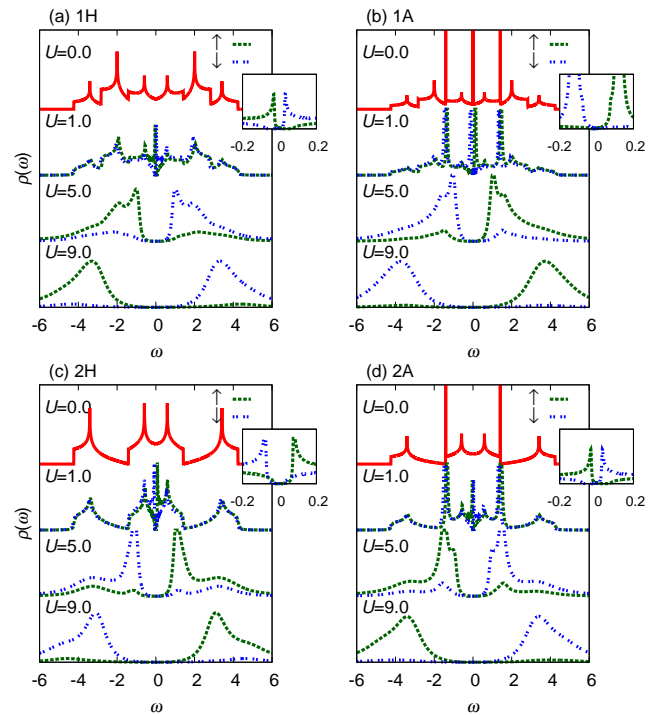}
\caption{(Color online) Local DOS $\rho(\omega)$ of (a) site H [(b) A] on the first layer and (c) site H [(d) A] on the second layer in the three-layer system ($L=3$) with $t_z=1.0$. Insets show $\rho(\omega)$ with $U=1.0$ near the Fermi energy, where vertical ranges are as the same as in the large panel. For comparison, the DOS at $U=0.0$ is extracted from Fig. \ref{fig_ggdos}(c). Note that relations $\rho_{mA\sigma}(\omega)=\rho_{mB\sigma}(\omega)$ ($m=1,2,3$) and $\rho_{1\gamma\sigma}(\omega)=\rho_{3\gamma\sigma}(\omega)$ ($\gamma=H,A,B$) exist.}
\label{fig_dos_3lay}
\end{center}
\end{figure}
%%%%%%%%%%%%%%%%%%%%%%%%%%%%

Next, we discuss the magnetism in the odd-$L$-layer lattices on the basis of the DMFT calculations explained in Sec.\,\ref{sec_method}.
The noninteracting DOS of these lattices have the flat band at $\omega=0$ as discussed in Sec.\,\ref{sec_dos}.
In particular, we first pick up the three-layer lattice as an typical example.
We set $t_z=t=1.0$. %and consider half filling.

Figure\,\ref{fig_mag_3lay} shows sublattice (local) magnetization $M_{m\gamma}$, where $M_{1\gamma}=M_{3\gamma}$ and $M_{mA}=M_{mB}$ because of the mirror symmetries.
% (m=1,2,3$ and $\gamma=\text{H,A,B})$
% through $x=y$ and $z=0$ planes, respectively.
For the magnetization on site $m\gamma=1A$, we find characteristic behaviour as follows: Magnetization $M_{1A}$ suddenly becomes a finite value of $-0.125$ with the infinitesimal interaction ($\delta U$), and then its magnitude continuously increases with increasing $U$, whereas $M_{1H}$ starts from zero and gradually grows as $U$ increases.
The singular behaviour, namely, a jump in $M_{1A}$ at $\delta U$, signals the emergence of the ferromagnetic state.
In contrast, the second layer ($m=2$) shows no characteristic behaviour: $M_{2\gamma}$ gradually changes in the same manner as $M_{1H}$.

The $m\gamma$-dependent magnetization processes result from the difference in the flat-band structures of the noninteracting DOS as mentioned in Sec.\,\ref{sec_dos}.
Because of the instability resulting from the flat band at the Fermi energy, ferromagnetic ordering immediately occurs with the infinitesimal interaction $\delta U$ in the first layer, which leads to a jump in $M_{1A}$ at $\delta U$.
On the other hand, all sites in the second layer and also site $H$ in the first layer have no flat bands as shown in Fig.\,\ref{fig_ggdos}\,(c).
The magnetization on these sites is caused by the antiferromagnetic correlation between adjacent sites. %along the $z$ direction.
Thus, the magnitude of $M_{2\gamma}$ ($M_{mH}$) gradually increases with a sign opposite to $M_{1\gamma}$ ($M_{mA}$).
The magnetic ordering for $m\gamma=1A$ caused by the instability of the flat band triggers inter- and intra-layer antiferromagnetic correlations.

\if0
Such indirect correlations are effectively taken into account through the DMFT self-consistent loop.
Namely, equation (7) describes that the global change of $\hat{G}_{{\bf k}\sigma}$ is caused by the local spin-dependent self-energies $\Sigma_{1A\uparrow}\not=\Sigma_{1A\downarrow}$ resulting from the local magnetization on the first layer; As a result, the Green's functions $G_{m\gamma\sigma}$ and ${\cal g}_{m\gamma\sigma}$ and the effective impurity problems for all $m$ and $\gamma$ can show spin dependence.
\fi
%The global change of the band structures is another key to the magnetism in this system.
%Thus, we obtain the flat band ferromagnetism only on the surface layers, which means that different crossover behaviour appear on the surface and second layer. These different magnetization processes trigger the novel layer magnetization flat band ferromagnetic behaviour discussed in the next paragraph.

%}% end green

%{\color{green}
To further discuss the above magnetization process through the dynamical quantities, we calculate the local DOS by changing interaction $U$ from weak to strong.
We show the results for $U=1.0$, $5.0$, and $9.0$ in Fig.\,\ref{fig_dos_3lay},  and we also show the noninteracting DOS for comparison,
where $\rho_{1\gamma\sigma}(\omega)=\rho_{3\gamma\sigma}(\omega)$ and $\rho_{mA\sigma}(\omega)=\rho_{mB\sigma}(\omega)$ because of the mirror symmetries.

We first focus on the DOS for the weak interaction $U=1.0$, which are shown in the second spectra from the top in Fig.\,\ref{fig_dos_3lay}\,(a)-(d).
By comparing these DOSs with those for $U=0$, we find that $\rho_{1A}(\omega)$ for $U=1.0$ shows split flat bands at around the Fermi energy, which is clear evidence of the emergence of the ferromagnetic ordering.
Importantly, the flat bands away from the Fermi energy are hardly split at all, suggesting that only one flat band just at the Fermi energy can be regarded as the origin of the flat-band instability toward magnetism.
We find that the other DOS for $U=1.0$ also show the gapped structures at the Fermi energy.
However, note that these gap structures caused by another mechanism different from the flat-band instability.
In fact, as shown in the insets, all gapped structures of the DOS except for $\rho_{1A}(\omega)$ show van Hove singularities at the edges of the gaps.
The appearance of such band structures after the magnetic transition results from the change of the band structure (band folding) due to the symmetry breaking caused by the antiferromagnetic ordering. %correlations between adjacent sites mentioned above.
%{\color{blue}
Note that for $L=1$ we do not observe this singularity at gap edges \cite{Noda2009} because of the Dirac-semi-metallic feature of the $\rho_{1H}(\omega)$ that has no DOS at the Fermi energy [see Fig.\,\ref{fig_ggdos} (a)]. 
%}% end of blue

In the insets of Fig.\,\ref{fig_dos_3lay}, we find another interesting feature: The gap energies of the DOSs for the sites labeled $m\gamma=1A$ and $2H$ are larger than those for $1H$ and $2A$.
The former (latter) sites are dominantly occupied $\sigma=\downarrow (\uparrow)$ atoms (cf.\,Fig.\,\ref{fig_mag_3lay}), and the total number of $\downarrow$-spin atoms are large as a result of the ferromagnetic ordering.
These facts imply that the majority atoms with $\sigma=\downarrow$ require a larger energy for the single particle excitation than the minorities.
This feature can be seen only in the weakly interacting region as mentioned below.

As $U$ increases, for $U=5.0$ and $9.0$,
the DOS for all sites loses the detailed structures and finally exhibit similar structures that recall upper and lower Hubbard bands.
%, which is described by ${\rm Im}\Sigma(\omega) \not = 0$ for $\omega$ away from the Fermi energy.
The change in the band structures indicates that the mechanism of the magnetic ordering changes from the flat-band picture to the localized Heisenberg picture, as discussed in the previous study \cite{Noda2009}.
The occurrence of this crossover can be confirmed from the magnetizations in Fig.\,\ref{fig_mag_3lay}, where $M_{m\gamma}$ for the strongly interacting region ($U\gtrsim 8$) shows saturation behaviour.
We note that, in contrast to the weakly interacting region, the DOS for the strongly interacting one shows no gap-energy difference depending on the site.
%Namely, the gap-energy difference can be seen only  for the weakly interacting region, where the band picture is good.

We should comment that the magnetism in the general-odd-$L$ layers can be understood from the simple extension of the above discussions.
This is because, qualitatively, the keys to the magnetism are the flat-band structures as mentioned in Sec.\,\ref{sec_dos}.
The quantitative difference will be discussed below in Sec.\,\ref{sec_L_dep}, which also addresses the important question how the magnetism is on the infinite-layer lattice ($L=\infty$).
%The key to this magnetism can be understood from the flat-band structures in the noninteracting DOSs.

%the following discussions to general odd $L$.

%}%end green

%%%%%%%%%%%%%%%%%%%%%%%%%%%%%
\subsection{Two-layer system as a typical example of even $L$}
\label{sec_even}
%%%%%%%%%%%%%%%%%%%%%%%%%%%%%

%%%%%%%%%%%%%%%%%%%%%%%%%%
\begin{figure}[tb]
\begin{center}
\includegraphics[clip,width=0.9\linewidth]{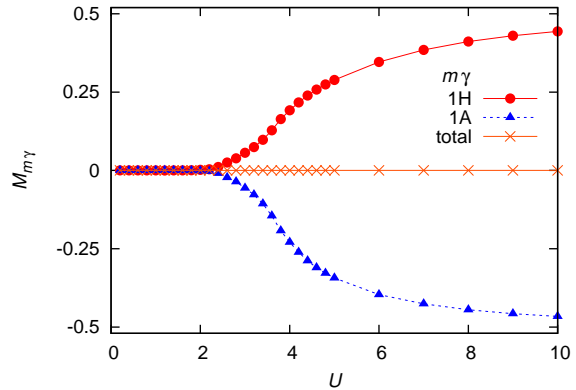}
\caption{(Color online) Sublattice magnetizations of site $\gamma(=H,A,B)$ on $m(=1,2)$-th layer in the two-layer system ($L=2$) with $t_z=1.0$. Only the first layer results are shown because of $M_{1\gamma}=-M_{2\gamma}$ and $M_{mA}=M_{mB}$}
\label{fig_mag_2lay}
\end{center}
\end{figure}
%%%%%%%%%%%%%%%%%%%%%%%%%%%%

%%%%%%%%%%%%%%%%%%%%%%%%%%
\begin{figure}[tb]
\begin{center}
\includegraphics[clip,width=1.0\linewidth]{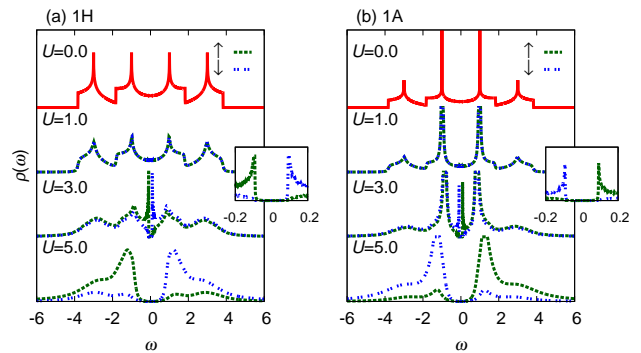}
\caption{(Color online) Local DOS of (a) site H [(b) A] on the first layer in the two-layer system ($L=2$) with $t_z=1.0$. For comparison, the DOS at $U=0.0$ is extracted from Fig. \ref{fig_ggdos}(b). Note that relations $\rho_{mA\sigma}(\omega)=\rho_{mB\sigma}(\omega)$ ($m=1,2$) and $\rho_{1\gamma\sigma}(\omega)=\rho_{2\gamma\overline{\sigma}}(\omega)$ ($\gamma=H,A,B$) exist.}
\label{fig_dos_2lay}
\end{center}
\end{figure}
%%%%%%%%%%%%%%%%%%%%%%%%%%%%

%{\color{green}
Next, we discuss the magnetism on the even-$L$-layer lattices, where the noninteracting DOS have no flat bands at the Fermi energy.
We investigate the two-layer $(L=2)$ lattice with $t_z=1.0$ as a typical example of them.
The following discussion can be easily extended to general even-$L$-layer lattices in the same manner as the odd-$L$ case.

Figure\,\ref{fig_mag_2lay} shows sublattice magnetizations $M_{m\gamma}$ for $L=2$, where $M_{1\gamma}=-M_{2\gamma}$ and $M_{mA}=M_{mB}$ because of the mirror symmetries.
The magnetization $M_{m\gamma}$ stays zero for the weakly interacting region, and it becomes finite at critical interaction $U_c \sim 2$, where a magnetic transition occurs.
The suppression of the magnetic ordering for $U<2$ is a consequence of the following facts:
Since the DOS for all $m\gamma$ have no flat bands at the Fermi energy as discussed in Sec.\,\ref{sec_dos},
there is no flat band instability toward magnetic ordering for small $U$.
In addition, we should note that, other kinds of instabilities, such as the nesting, also do not exist.
We find that the phase transition occurs when $U$ becomes comparable to a specific energy scale $\Delta$, i.e., the difference in the spectral positions of the flat bands $\Delta= 2t_z$ for $L=2$ [see Fig.\,\ref{fig_ggdos}\,(b) or Appendix].
This can be intuitively understood as follows;
an (inelastic) excitation caused by $U$ restores the instability of the two flat bands at $\omega=\pm\Delta/2$ under the condition $U\sim \Delta$.

The above points can be confirmed by the DOS shown in Fig.\,\ref{fig_dos_2lay}, where symmetries impose $\rho_{1\gamma\sigma}(\omega)=\rho_{2\gamma \bar{\sigma}}(\omega)$ and $\rho_{mA\sigma}(\omega)=\rho_{mB\sigma}(\omega)$, where $\bar{\sigma}$ is the opposite spin of $\sigma$.

For weak interaction $U=1.0$, we find no spectral gap, suggesting no magnetic ordering,
while for $U=3.0$, we find a gap accompanied with van Hove singularities, which  suggests the magnetic ordering with band folding.
We can thus conclude that at $U_c\sim 2$ this is a phase transition from a paramagnetic metal to an antiferromagnetic insulator.
Interestingly, the gap structure caused by the restored flat-band instability is analogous to those in the DOS for the sites having no flat bands for $L=3$ (see Fig.\,\ref{fig_dos_3lay}).

As $U$ further increases, the DOS becomes the upper and lower Hubbard bands, suggesting that the localized Heisenberg picture governs the magnetism.
%whose picture is naturally connected to the localized Heisenberg picture.
By comparing this result with that shown in Fig.\,\ref{fig_dos_3lay},  we find that the difference in the magnetism between even- and odd-$L$-lattices disappears in the Heisenberg limit.
This is because the band structure in the ${\bf k}$ space is irrelevant to the magnetism in this limit, whereas the bipartite structure in the ${\bf x}$ space is important.
The bipartite structure is the same for all $L$, while the flat-band structure at around the Fermi energy strongly depends on $L$.
Note that, even though the odd-even difference in the DOS disappears, the difference in the total magnetization ensured by the Lieb theorem, $M_{\rm tot}=0.5$ for odd $L$ and 0 for even $L$, is still satisfied due to the alternating ordering along the $z$ direction.

%}

%%%%%%%%%%%%%%%%%%%%%%%%%%%%%
\subsection{$L$ dependence and the $L=\infty$ limit}
%%%%%%%%%%%%%%%%%%%%%%%%%%%%%
\label{sec_L_dep}

%{\color{green}
%As discussed in the previous two sections, the magnetism shows strong even and odd $L$ dependence.

%%%%%%%%%%%%%%%%%%%%%%%%%%
\begin{figure}[tb]
\begin{center}
\includegraphics[clip,width=\linewidth]{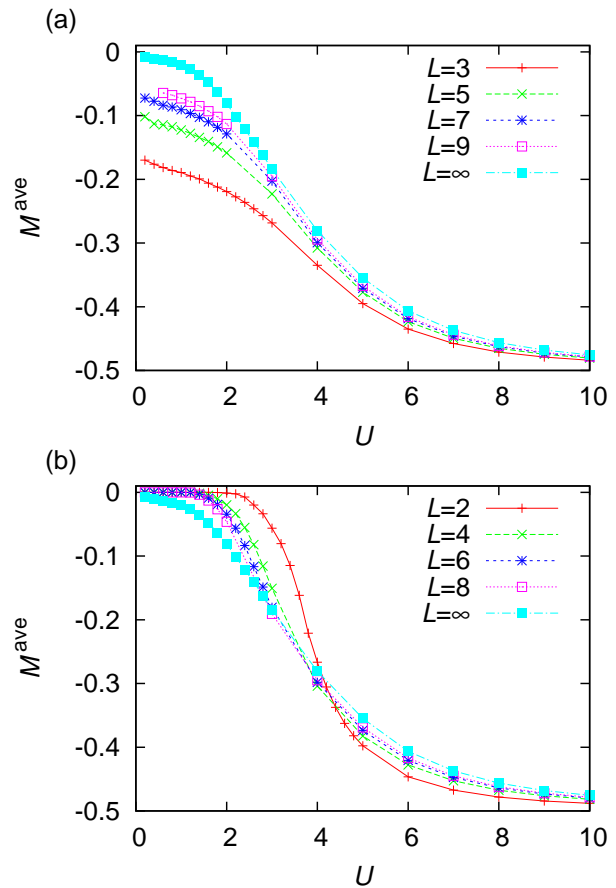}
\caption{(Color online) Average of magnetization as a function of $U$ with the condition $t_z=1.0$. (a) Odd-layer systems. (b) Even-layer systems. 
}
\label{fig_to3d}
\end{center}
\end{figure}
%%%%%%%%%%%%%%%%%%%%%%%%%%%%
%Finally, we discuss how the different behaviour between the even and odd layered systems disappears in the approach to the $L=\infty$.
We further investigate the $L$ dependences of magnetism, and we also address the infinite-layer lattice that corresponds to the three-dimensional layered Lieb lattice.
In Fig.\,\ref{fig_to3d}, we show the $L$-dependence of the $M^{\rm ave}$-$U$ curves, where $M^{\rm ave}$ is the average magnetization defined as
\begin{equation}
M^{\text{ave}}=-\sum_{\gamma=A, B, H}\sum_{m=1,\dots,L} |M^{}_{m\gamma}|/3L.
\end{equation}
Here, we add an overall minus sign in Eq. (9) to use the same sign of magnetization of sites where the flat-band ferromagnetic behaviour appears in $L=3$ (see site 1A in Fig. \ref{fig_mag_3lay}).
Figure\,\ref{fig_to3d}\,(a) shows that, for odd $L$, average magnetizations in the weak coupling region decrease as $L$ increases.
On the other hand, Fig.\,\ref{fig_to3d}\,(b) shows that, for even $L$, the critical interaction $U_c$ decreases as $L$ increases.
In the strongly interacting region, the $M^{\text{ave}}$-$U$ curves show the similar behaviour irrespective of whether $L$ is even or odd. 
We confirm the disappearance of the odd-even difference again, which we have already discussed in terms of the DOSs at the Heisenberg limit in the last paragraph in Sec. IV.C.
We also find that the curves for both odd- and even-$L$-layer lattices naturally approach those in the same limit of $L = \infty$ (the three-dimension limit).
%These asymptotic behaviour are discussed later.

As shown in Fig.\,\ref{fig_to3d}, in the three-dimensional layered Lieb lattice ($L=\infty$), we find that the magnetic ordering appears at infinitesimal $U$.
% due to the instability of the flat band at the Fermi energy.
The asymptotic behaviour of odd-$L$-layer lattices indicates that this magnetism for $L=\infty$ is a consequence of the flat-band instability.
\begin{comment}
On the other hand, it seems at first glance that the even-$L$-layer lattices without flat-band instability shows rather curious asymptotic behaviour as follows.
%{\color{blue}
Naively, we can expect that the three dimensional lattice ($L=2l=\infty$) will show a finite $U_c$,
%, because they have broadened band structures.
%; {\it e.g.}, the square lattice has the bandwidth $8t$, while the cubic lattice has the wider bandwidth $12t$.
because the flat band becomes dispersive with a broadening of $4t_z$ as mentioned in Sec.\,\ref{sec_dos} (and also in Appendix). 
%} % end of blue
However, interestingly, the critical interaction $U_c$ decreases with increasing $L$. % that broadens flat bands.
This behaviour can be understood from the following striking feature: the flat-band instability is restored with infinitesimal $U$ at $l=\infty$ with $L=2l$.
%, because  $\Delta$, the energy difference of two flat bands nearest to the Fermi energy, becomes zero asymptotically (see Appendix).
%This feature is also regarded as the origin of the characteristic even-odd dependence.
%Thus, the different behaviour between odd and even layered systems become naturally identical in the infinite layer limit.
\end{comment}
On the other hand, in the even-$L$ layer lattices without flat-band instability, the curious asymptotic behaviour (i.e., the decrease in $U_c$ with increasing $L$)  reflects the striking feature: the flat-band instability is restored with infinitesimal $U$ at $l = \infty$ with $L = 2l$. This completely contradicts our naive expectation that the three-dimensional lattice ($L= \infty$) will show a finite critical $U_c$, because the flat band becomes dispersive with a broadening of $4t_z$ as mentioned in Sec.\,\ref{sec_dos} (and also in Appendix).

%%%%%%%%%%%%%%%%%%%%%%%%%%
\begin{figure}[tb]
\begin{center}
\includegraphics[clip,width=0.9\linewidth]{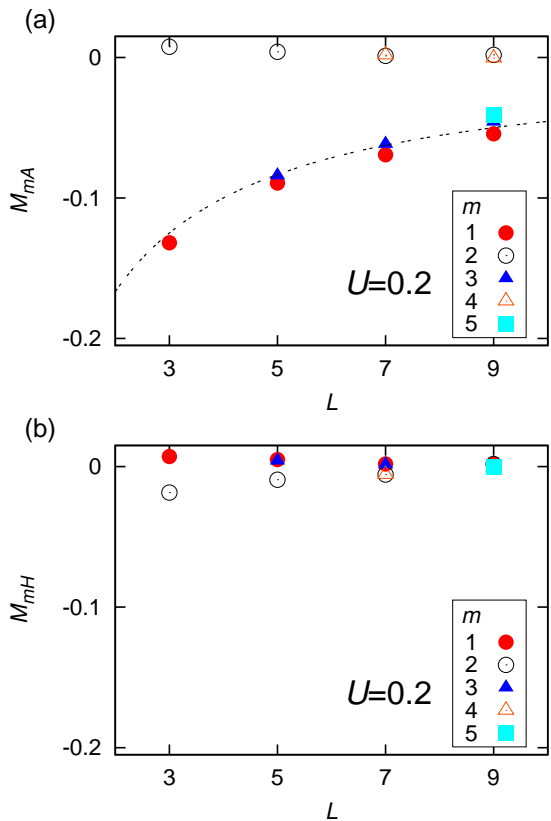}
\caption{(Color online) Magnetization of site A (a) and site H (b) in $L=3,5,7,9$ for $U=0.2$. The dashed line shows $-0.5/(L+1)$, which is equal to $-0.25/l$ in the $L=2l-1$ layered system.}
\label{fig_odd}
\end{center}
\end{figure}
%%%%%%%%%%%%%%%%%%%%%%%%%%%%
We discuss the asymptotic behaviour with increasing $L$ in more detail.
We first show the results for the odd-$L$-layer lattices and %, where the flat-band magnetism appears at infinitesimal $U(=\delta U$).
focus on a characteristic quantity $M^{\delta U}_{mA}$, the magnetization at infinitesimal $\delta U$.
%As shown in Fig. 2 (b) in Sec. B, $M^{\delta U}_{1A}=0.125$ for $L=3$.
%Our previous study for $L=1$ two-dimensional Lieb lattice clarify that $M^{\delta U}_{1A}$ is 0.25.
%We also calculate $M^{\delta U}_{1A}$ by changing the natural number $l$ with $L=2l-1$.
Figure\,\ref{fig_odd} shows  $M^{\delta U}_{m\gamma}$ calculated by changing the natural number $l$ with $L=2l-1$, where we set $\delta U$ to a rather large value of $0.2$ because of the numerical limitation.
We find $M^{\delta U}_{mA}$ for odd $m$ is proportional to $1/l$, while $M^{\delta U}_{mA}$ for even $m$ and $M^{\delta U}_{mH}$ for all $m$ stay zero (within the present numerical precision).
This difference between even and odd $m$ is attributed to the noninteracting DOS.
We can simply explain the origin of $\propto 1/l$ behaviour for odd $m$ as follows:
The $L$-layer lattices have  $3\times L$ multibands, and
only one of the $3\times L$ bands, that is, the flat band at the Fermi energy, takes part in the ferromagnetism with infinitesimal $U$.
This means that the weight of the flat-band related to the ferromagnetism decreases with increasing $l$ as $\propto 1/l$.
The dotted line shows the asymptotic behaviour $1/4l$, which is evaluated analytically beyond the above intuitive discussions (see Appendix).
%More detailed discussion about this asymptotic behaviour is left for Appendix.
The Lieb theorem provides the same conclusion: the difference in the number of sublattices $N_\alpha-N_\beta$ stays one as $L$ increases even though the unit cell size increases as $3\times L$.

We also explain the asymptotic behaviour in the even-$L$-layer lattices.
The characteristic of even $L$ lattices is the finite critical interaction $U_c$.
As mentioned in Sec.\,\ref{sec_even}, the magnetism occurs when the strength of interaction $U$ becomes comparable to $\Delta$ the energy difference in the spectral positions of the two flat bands nearest to the Fermi energy, where the flat-band instability is restored as discussed in Sec.\,\ref{sec_even}.
As derived in Appendix, we find $\Delta \propto 1/l$, which determines the asymptotic behaviour of the even-$L$-layer lattices shown in Fig.\,\ref{fig_to3d}\,(b).
This clearly shows that the flat-band instability is restored with infinitesimal $U$ at $L=\infty$, and the difference between even and odd $L$ naturally disappears at $L=\infty$.
%We confirm this below.

%}%end green

%%%%%%%%%%%%%%%%%%%%%%%%%%%%%
\subsection{$t_z$ dependence and the $t_z=0$ limit}
\label{sec_approach}
\label{sec_tz_dep}
%%%%%%%%%%%%%%%%%%%%%%%%%%%%%

%%%%%%%%%%%%%%%%%%%%%%%%%%
\begin{figure}[tb]
\begin{center}
\includegraphics[clip,width=0.9\linewidth]{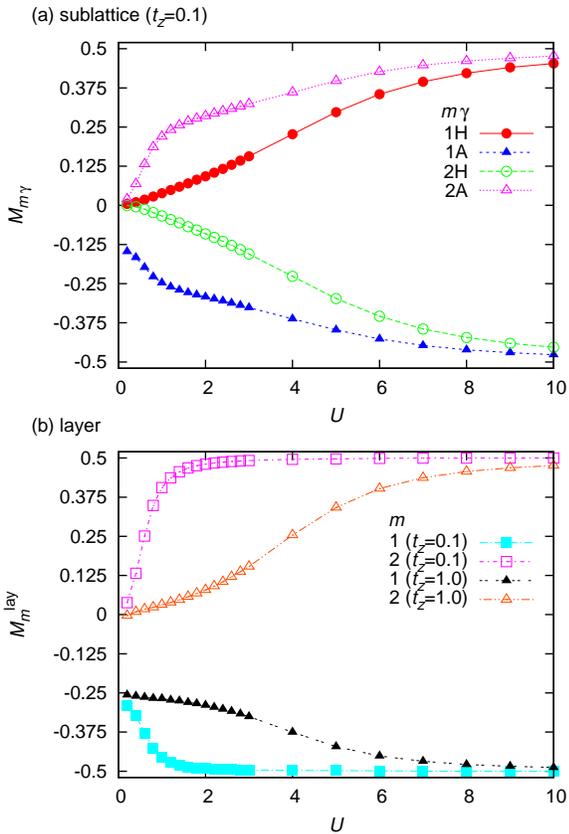}
\caption{(Color online) Magnetization as a function of $U$ with the condition $t_z=0.1$ for $L=3$. (a) Sublattice magnetization. (b) Layer magnetization. In the lower figure, results for $t_z = 1.0$ are also shown for comparison. Note that relations $M_{mA}=M_{mB}$ ($m=1,2,3$), $M_{1\gamma}=M_{3\gamma}$ ($\gamma=H,A,B$) and $M^{\text{lay}}_{1}=M^{\text{lay}}_{3}$ exist.}
\label{fig_mag_tz}
\end{center}
\end{figure}
%%%%%%%%%%%%%%%%%%%%%%%%%%%%

%{\color{green}
We finally investigate the $t_z$ dependence of the magnetism by changing $t_z$ toward $\sim 0$.
In the limit of $t_z \rightarrow 0$, the present system for all $L$ becomes equivalent to the two-dimensional lattice that has already been investigated in Ref. \cite{Noda2009}.
We thus perform the calculations on the three-layer lattice for small $t_z=0.1$ and compare these results with those for $t_z=1.0$ already shown in Fig.\,\ref{fig_mag_3lay}.

Figure\,\ref{fig_mag_tz}\,(a) shows the sublattice magnetization $M_{m\gamma}$ for $t_z=0.1$ as a function of $U$.
We find that $M_{m\gamma}$ for $\gamma=A$ steeply grows for $U<1.0$, and then it gradually grows for $U> 1.0$ and shows saturation at around $U\sim 8.0$.
On the other hand, $M_{m\gamma}$ for $\gamma=H$ shows no characteristic behaviour at around $U=1.0$ and shows saturation similar to the above at around $U\sim 8.0$.
The difference between $M_{m\gamma}$ for $\gamma=A$ and $H$ sites is attributed to the flat bands that appear only in $\rho_{mA}(\omega)$.
The saturations for both $\gamma=A$ and $H$ suggest the crossover from the flat band to the Heisenberg magnetisms as already mentioned above.
For small $t_z=0.1$, at around $U=1.0$, we observe the characteristic behaviour not observed in $M_{m\gamma}$ for $t_z=1.0$ (see Fig.\,\ref{fig_mag_3lay}).
%how properties of multilayered lattices approach to the two dimensional Lieb lattice () and the infinite layer ($L \rightarrow \infty$).

%For this purpose,
To clarify the difference between the results for smaller and larger $t_z$, we also show in Fig.\,\ref{fig_mag_tz}\,(b) the layer magnetization defined as $M^{\text{lay}}_{m}=\sum_{\gamma=H,A,B}M_{m\gamma}$, which characterizes the (macroscopic) population imbalance between the $\sigma=\uparrow$ and $\downarrow$ atoms in the corresponding layer $m$.
%a sum of sublattice magnetizations on $m$th layer.
Note that, with the infinitesimal interaction, this layer magnetization also shows a clear jump, which is a feature of the flat-band ferromagnetic state.
We find that $M^{\text{lay}}_{m}$ shows saturation behaviour at around $U=1.0$.
This suggests that the layer magnetization clearly signals the crossover.
The saturated layer magnetization shows antiferromagnetic alternating behaviour: $M^{\text{lay}}_{m}=(-1)^m 0.5$.
Note that, in the two-dimensional Lieb lattice ($t_z=0$), the layer magnetization always stays at 0.5 when $U$ is finite, which is a manifestation of the Lieb theorem.
From this point, we can understand the origin of
%the behaviour $U>1.0$ has analogy to that in the two-dimensional Lieb lattice.
the crossover at around $U=1.0$ as follows:
for $U<1.0$, only one of the flat bands at the Fermi energy takes part in the flat-band magnetism as discussed above (see Fig.\,\ref{fig_dos_3lay} and related discussions in Sec.\,\ref{sec_odd}).
In contrast,  for $U>1.0$,
all of the flat bands join the magnetic ordering, even though they are not at the Fermi energy.
The latter situation can be regarded as analogous to the flat-band magnetism seen in the ensemble of two-dimensional Lieb lattices.
We should note that the interlayer hopping still causes antiferromagnetic correlations between the Lieb lattices.

We can thus conclude that the crossover at $U=1.0$ for $t_z=0.1$ originates from the change in the picture of the magnetism from that in the quasi-three-dimensional layered Lieb lattice to that in the correlated ensemble of the two-dimensional Lieb lattice.
We can now evaluate that the crossover occurs when $U$ becomes much larger than $W=2\sqrt{2}t_z$, which corresponds to the energy difference between the two flat bands farthest from the Fermi energy (see Appendix).
For $t_z=1.0$ (and also larger $t_z$), this crossover cannot be seen, because it is covered by the flat-band to Heisenberg crossover that occurs when $U$ becomes comparable to the bandwidth of the two-dimensional Lieb lattice $4\sqrt{2}t(\sim 6)$.
In the Heisenberg limit, the flat-band structure is irrelevant to the magnetism, which can be confirmed from the fact that the DOS for strong $U$ shows no band structures except for the upper and lower Hubbard bands  as mentioned above and also in Ref. \cite{Noda2009}.

%}%end color blue

% This value is corresponding to the total magnetization in the two dimensional Lieb lattice when $t_z \rightarrow 0$. In this sense, $M^{\text{lay}}=0.5$ means that multilayer recover the magnetizaion of the two dimensional Lieb lattice. We infer that, when all flat bands participate into the magnetization, the system shows $M^{\text{lay}}=0.5$. This fact is a generalization of the two dimensional Lieb lattice, where the flat band participates magnetization with infinitesimal interaction and induces $M^{\text{lay}}=0.5$.

%This novel layer magnetization behaivour is also understood as a result of the different magnetization process on layers.

%

%Before our conclusion, we give two comments about experiment. In our study, we reveal that not only sublattice but also layer magnetizations show the flat band ferromagnetic behaviour. This is a striking feature of the multilayered Lieb lattice. We also mention about an advantage of making finite layer systems using addtional lasers. In this situation, we expect that we can use a path to anisotropic hopping to the $z$ direction, gain an advantage to observe the magnetization. In the ETH experiment, they use this technique and observe the short-range mangetic correlations \cite{Greif2013}.

%%%%%%%%%%%%%%%%%%%%%%%%%%%%%
\section{conclusions}
%%%%%%%%%%%%%%%%%%%%%%%%%%%%%
\label{sec_con}

We investigate the magnetic properties of two-component fermions in a multilayer Lieb lattice at half filling and at zero temperature using the DMFT combined with the NRG. 
%We elucidate that magnetization behaviour is different in even or odd layers. We reveal that the flat band ferromagnetic state only appears in odd-$L$ layers. For even-$L$ layers, when the interaction restores the flat band instability toward the antiferromagnetic ordering, the magnetic transition occurs. This even-odd difference disappears for $L=\infty$. We reveal that, in all multilayers, magnetic properties in the weak interaction region are dominated by flat bands, while that in the strong interaction region can be understood by the Heisenberg spin picture. We elucidate the interlayer correlation effects, which induce another crossover and the antiferromagnetic correlations between layers.
We elucidate that even- and odd-$L$ layers show different magnetization behaviours. The flat-band ferromagnetic state appears only in odd-$L$ layers. For even-$L$ layers, the transition occurs when the interaction restores the flat-band instability toward the antiferromagnetic ordering. This odd-even difference disappears in the limit of $L=\infty$. On the other hand, as common features seen in all layers, the magnetic properties in the weak interaction region are dominated by flat bands, while those in the strong interaction region can be well understood by the Heisenberg spin picture. We further elucidate the interlayer correlation effects, which induce another crossover and the antiferromagnetic correlations between layers.

%{\color{blue}
In this paper, we restrict our calculations to zero temperature. It is also important to briefly discuss the property at finite temperatures. 
We expect that we can observe finite transition temperatures in the present multilayer Lieb lattices thanks to the weak three-dimensional effects, 
although the Marmin-Wagner theorem theoretically prohibits the transition in such quasi-two dimensional systems. 
From the numerical results, we have revealed the crossover from the flat-band to the Heisenberg picture with increase in the interaction strength. 
This clearly indicates that the transition temperature is the highest in the crossover region by noting the same mechanism appearing in the cubic lattice \cite{Pruschke2005}. 
An interesting future work is to extend our theory to the case at finite temperatures. 
We will address this problem and publish it elsewhere.

We mention experimental advantages of making the multilayer structure. 
%First, we can use layer magnetizations to detect flat band ferromagnetic state. This means that it is enough to observe number difference between layers instead of addressing sublattice magnetization. 
First, we can detect the flat-band ferromagnetic state by only measuring layer magnetizations without addressing sublattice magnetization as shown in Fig.\,\ref{fig_mag_tz} (b). 
This is a novel property of the multilayer Lieb lattice. Second, magnetic correlations in a multilayer system can be enhanced by the following experimental technique. 
%In the ETH group \cite{Greif2013}, an adiabatic change from the cubic lattice to the anisotropic or dimerized cubic lattice causes local redistribution of entropy, which induces the enhancement of the short-range magnetic correlation. They succesfully observe this correlation. For multilayers, when the additional lasers along the $z$ direction are introduced, anisotropic interlayer hopping is addiabatically induced. This entropy redistribution enhances the magnetic correlation inside the multilayer system. This may help us to observe magnetic states by current experimental techniques. These facts encourage us to consider the multilayer lattices as good candidates to observe the flat band ferromagnetic state. 
In Ref. \cite{Greif2013}, the ETH group adiabatically changed the cubic lattice to an anisotropic or dimerized cubic lattice. 
This process causes local redistribution of entropy, leading to a great enhancement of the short-range magnetic correlation. They successfully observed the enhanced magnetic correlation. For multilayer Lieb lattices, we can also  adiabatically  induce anisotropic interlayer hopping by applying additional lasers in the $z$ direction. The  entropy redistribution similarly enhances the magnetic correlation inside the multilayer system. 
%The multi layer Lieb lattice is the promising candidate for observing the flat band ferromagnetic state.
These two advantages encourage us to consider the multilayer lattices as promising candidates for observing the flat-band ferromagnetic state.

\section{acknowledgements}
%%%%%%%%%%%%%%%%%%%%%%%%%%%%%

We are grateful to N. Kawakami and Y. Takahashi for valuable discussions. This work was partly supported by Grants-in-Aid for Scientific Research (No. 25287104).

%\end{acknowledgments}

%%%%%%%%%%%
\appendix*
\section{}
%{\color{blue}
This appendix shows simple analytical discussions about the local DOS of the noninteracting atoms in the $L$-layer Lieb lattices with an open boundary condition along the $z$ direction.
We calculate the spectral positions and weights of flat bands in the noninteracting local DOS, and
we simply explain why the difference between odd and even $L$ appears.
%, and also what determines the magnitudes of the onset magnetization for odd $L$ and the critical interaction strength $U$ for even $L$.
We also address what determines the asymptotic behaviour from the finite-$L$-layer to the three-dimensional (infinite) layer Lieb lattices for both even and odd $L$.

Here, we discuss the noninteracting local DOS $\rho_{L,m\gamma}(\omega)$ for $L=1, 2, 3, \cdots$ with $\gamma=A, B, H$ and $m=1, 2, \cdots, L$.
The DOS for $L=1$ is of the well-known two-dimensional Lieb lattice [see Fig.\,\ref{fig_ggdos}\,(a)].
It is convenient to use the following expression for them: $\rho^{\rm 2D}_{\gamma}(\omega) [\equiv \rho_{1,1\gamma}(\omega)]$.
The DOS for multilayer lattices with $L \ge 2$ can be obtained by the simple extension from the two-dimensional case because the noninteracting Hamiltonian, ${\cal H}_{\rm Lieb}+{\cal H}_{z}$, can be solved by variable separation.
Namely, $\rho_{L,m\gamma}(\omega)$ is written as follows:
\begin{equation}
\rho_{L,m\gamma}(\omega)=\sum_{n=1}^{L} |u_{m,n}|^2 \rho_{\gamma}^{\rm 2D}(\omega-\lambda_{n}),
\end{equation}
where $\lambda_{n}$ and $u_{m,n}$ are eigenvalues and corresponding eigenvectors of the following matrix $\hat{H}'_{z}$ with a dimension of $L\times L$:
\begin{equation}
\hat{H}'_{z}=
\begin{pmatrix}
0 \ & t_z &   & & \vspace{2pt}\\
t_z \ & 0 & t_z & & \\
 & t_z & \ddots & \ddots & \\
 &  & \ddots & \ddots & t_z \\
 & & & t_z & 0
\end{pmatrix}
.
\end{equation}
This matrix is the submatrix of $\hat{H}_{z}$ defined in the main text.
%, and the eigenvalues $\lambda_n$ describe the energy dispersion as a function of $k_z$.
Figure\,\ref{fig_ggdos} confirms the above discussion.
%{\color{blue}
This matrix can be solved analytically, which is shown in detail with our physical interpretation in the following several paragraphs. Those who are familiar with this solution can skip those paragraphs.
%}

In what follows, for simplicity, we only focus on the flat bands and neglect the other DOS structures.
This simplification is very rough but well describes the physics of the flat-band magnetism that occurs at the infinitesimal $U$.
This simplification yields $\rho^{\rm 2D}_{A}(\omega)=\rho^{\rm 2D}_{B}(\omega)=(1/2) \delta(\omega)$ and $\rho^{\rm 2D}_{H}(\omega)=0$.
The positions (weights) of the flat bands in $\rho_{L, mA(B)}(\omega)$ are determined from  $\lambda_n$ ($|u_{m,n}|^2$).
Note that the sum rule $\sum_{m\gamma}\int d\omega \rho_{L, m\gamma}(\omega)=L$ should be satisfied because we now consider only $L$ flat bands and neglect the other $2L$ bands.

The eigenvalues $\lambda_m$ in energy units of $t_z$ can be calculated from the following recurrence relation:
\begin{equation}
\label{eq_eval_rec}
\begin{split}
p_k(\lambda)&=\lambda p_{k-1}(\lambda)-p_{k-2}(\lambda) \, (k=3,..,L), \\
p_2(\lambda)&=\lambda^2-1, {\ \ \ \text{and}\ \ \ } p_1(\lambda)=\lambda, 
\end{split}
\end{equation}
where $p_L(\lambda)=0$ is equivalent to $|\lambda \hat{I} - \hat{H}'_{z}|=0$ and the eigenvalues $\lambda_n$ satisfy $p_L(\lambda_n)=0$.
On the other hand, the eigenvectors ${\bf u}_n=(u_{1, n}, u_{2, n}, \cdots, u_{L, n})$ can be calculated from $(\lambda_n \hat{I}-\hat{H}'_{z}) {\bf u}_n = {\bf 0}$, which can be rewritten as
\begin{equation}
u_{m+1,n}=\lambda_n u_{m,n}-u_{m-1,n}, (m=1,2,3, \cdots, L-1),
\label{eq_evec_rec}
\end{equation}
where $u_{0,n}=0$ and $u_{1,n}=1$.
The obtained eigenvectors should be normalized as $\sum_m  |u_{m,n}|^2 =1$.

Equation\,(\ref{eq_eval_rec}) clearly explains why the difference between even and odd $L$ layers appears.
With $\lambda=0$, Eq. (\ref{eq_eval_rec}) reduces to $p_L(0)=p_{L-2}(0)$ with $p_{1}(0)=0$ and $p_{2}(0)\not = 0$.
This means that the flat band at the Fermi energy appears only for odd $L$.

From Eq.\,(\ref{eq_evec_rec}),
we can discuss the asymptotic behaviour of $M^{\delta U}_{m\gamma}$, the magnetization at infinitesimal interaction $\delta U$, for odd $L=2l -1$ with $l$ a natural number.
Substituting $\lambda_n=0$ into Eq.\,(\ref{eq_evec_rec}), we obtain the corresponding renormalized eigenvector as ${\bf u}_n= 1/\sqrt{l} \big(1, 0, -1, 0, 1, 0, \cdots, (-1)^{l+1}\big)$, which means that the weights of the flat band at the Fermi energy in local DOS are $1/l$ for odd-number-th layer and 0 for even-number-th layer.
This clearly shows that the flat-band magnetism appears only in the odd-number-th layers.
% with asymptotic behaviour of $M_{m\gamma}\propto 1/l$, .
%, which can be confirmed by Fig. 3 (a) for $l=1$.
The weights of the flat bands determine that $M^{\delta U}_{m\gamma}$ for odd $m$ asymptotically decreases with $1/4l$, where the factor of $1/4(=1/2\times 1/2)$ results from $1/2$ in the definition of $M_{m\gamma}$ and $1/2$ in $\rho^{\rm 2D}_{mA(B)}$ shown above while that for even $m$ stays $0$.
These are numerically confirmed as shown in Fig.\,\ref{fig_odd} and discussed in Sec.\,\ref{sec_L_dep}.

Furthermore, we can discuss the critical interaction strength $U$ for even $L(=2l)$.
As discussed in Sec. \ref{sec_even}, we assume  that the critical $U$ can be determined from the energy difference $\Delta$ between the two flatbands nearest the Fermi energy.
From Eq.\,(\ref{eq_eval_rec}), we can derive the explicit representation of the eigenvalues as $\lambda_n=-2 t_z \cos(n \pi/(L+1)), (n=1,2,\cdots,L)$.
We obtain $\Delta=\lambda_{l+1}-\lambda_{l}=4 t_z \sin(\pi/(2+4l)) = \pi t_z /l+{\cal O}(1/l^2)$ for large $L=2l$.
This means that the critical interaction strength $U$ also shows ($1/l$)-asymptotic behaviour as discussed in Sec.\,\ref{sec_L_dep}.
We should comment that neglecting the DOS except for the flat band as mentioned above is actually an oversimplification due to the finite (not infinitesimal) critical interaction.
Nevertheless, our numerical results with $t_z=1$ show a good agreement with the above simple analysis.
For example, as shown in Fig.\,\ref{fig_mag_2lay}, the critical $U$ for $L=2$ is about $\Delta \sim 2$.
%}%end color blue

%{\color{green}
We mention another energy scale $W=\lambda_{L}-\lambda_{1}$, which represents the energy difference between the two flat bands farthest from the Fermi energy.
We can rewrite $W=4 t_z \sin\big(\pi/2\times (L-1)/(L+1) \big)$, which reduces to $4t_z +{\cal O}(1/L^2)$ for a large $L$.
Note that $W$ can be regarded as the bandwidth of broadened flat bands in the limit of $L=\infty$, and the total bandwidth of the three bands in this three-dimensional limit is given by $4\sqrt{2}t+4t_z$.
This energy scale $W$ determines the characteristic crossover discussed in Sec.\,\ref{sec_tz_dep}.
%}%end color green

%Due to the numerical errors, however, we cannot precisely compare numerical calculations with the analysis.

%From the above simple analyses, we can explain the asymptotic behaviour from the finite layer to the three dimensional (infinite) layer Lieb lattices for both even and odd $L$.
%Here, we can explain at least quantatively the decreasing behaviour of the onset interaction with increasing the number of layer from this $1/l$, whereas our numerical results not match the behaviour due to the interaction being finite.

%\bibliography{multilieb.bib}

%%%%%%%%%%%%%%%%%%%%%%%%%%%%%

\end{document}